\documentclass[a4paper,11pt]{article}
\usepackage{pos}

\title{Impacts of the external environment on the Virgo detector during the third Observing Run}

\author*[a,b]{Nicolas Arnaud}

\affiliation[a]{ Laboratoire de physique des deux infinis Irène Joliot-Curie (IJCLab),\\
Universit\'{e} Paris-Saclay and CNRS/IN2P3, 91405 Orsay, France}

\affiliation[b]{European Gravitational Observatory (EGO),\\
I-56021 Cascina (PI), Italy}

\emailAdd{nicolas.arnaud@ijclab.in2p3.fr}

\abstract{Sources of geophysical noise (such as wind, sea waves and earthquakes) or of anthropogenic noise (nearby activities, road traffic, etc.) impact ground-based gravitational-wave (GW) interferometric detectors, causing transient sensitivity worsening and gaps in data taking. During the one year-long third Observing Run (O3: from April 01, 2019 to March 27, 2020), the Virgo Collaboration collected a large dataset, which has been used to study the response of the Advanced Virgo detector to a variety of environmental conditions. We correlated environmental parameters to global detector performance, such as observation range (the live distance up to which a given GW source could be detected), duty cycle and control losses (losses of the global working point, the instrument configuration needed to observe the cosmos). Where possible, we identified weaknesses in the detector that will be used to elaborate strategies in order to improve Virgo robustness against external disturbances for the next data taking period, O4, currently planned to start in March 2023. The lessons learned could also provide useful insights for the design of the next generation of ground-based interferometers.
}

\FullConference{%
  41st International Conference on High Energy physics - ICHEP2022\\
  6-13 July, 2022\\
  Bologna, Italy
}

\begin{document}
\maketitle

\section{Introduction}

The Advanced Virgo Detector~\cite{AdVPlus} (AdV) is a member of the global network of ground-based interferometers which has detected 90 gravitational-wave (GW) signals since 2015, published in the third issue of its catalog of transient sources: GWTC-3~\cite{2111.03606}. As the Observing Run 4 (O4) is about to start in a few months (after long delays mainly due to the covid-19 pandemic, which has impacted heavily all long-planned upgrades), the network includes four instruments: AdV, the two Advanced LIGO detectors~\cite{PhysRevD.102.062003}~-- located in Hanford (WA, USA) and Livingston (LA, USA)~-- and KAGRA~\cite{10.1093/ptep/ptab018}, built underground in Japan.
The previous Observing Run, O3, lasted almost one full year: O3a, from April 1st, 2019 to October 1st, 2019; a one-month commissioning break; 
O3b, from November 1st, 2019 to March 27, 2020, shortened by a month because 
of the first covid-19 lock-down period. 
During O3, the first long run of AdV, a large dataset was collected. In addition to contributing to the detection of the GW signals mentioned above
, it allowed the Virgo Collaboration to study the performance of the instrument, and to measure how various environmental noises impacted it~\cite{2203.04014}. 

\begin{figure}[!htbp]
    \begin{center}
      \includegraphics[width=0.45\columnwidth]{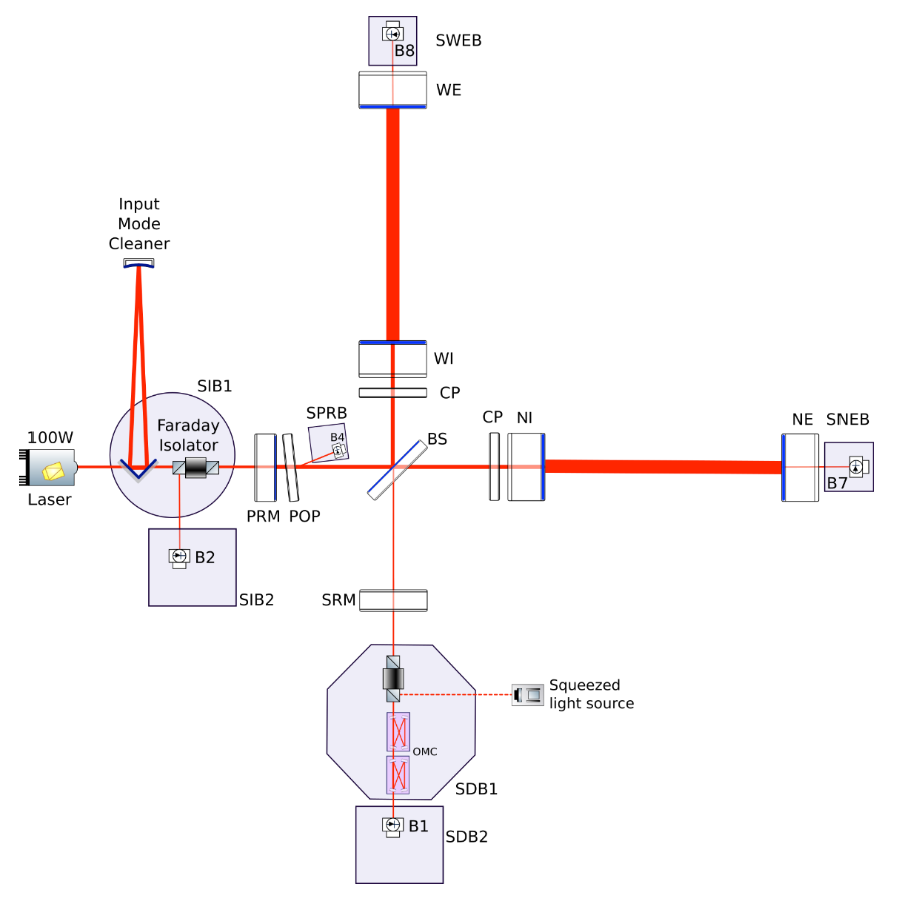}
      \includegraphics[width=0.53\columnwidth]{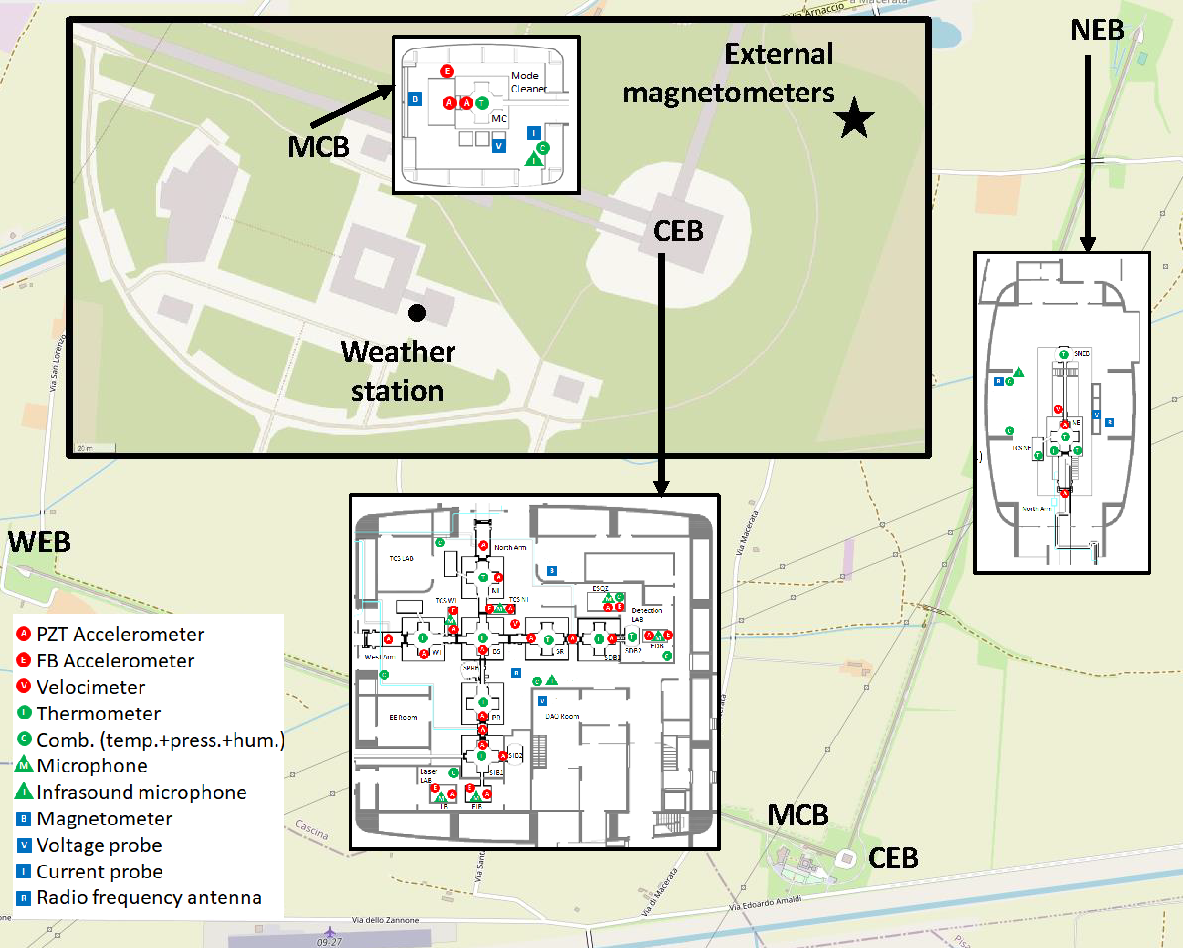}
    \end{center}
    \caption{Left: layout of AdV during the O3 run. Right: environmental monitoring during O3.}
    \label{fig:Virgo_O3}
\end{figure}

\section{The Advanced Virgo detector}

AdV is a power-recycled Michelson interferometer, with Fabry-Perot cavities in its 3-km long arms. Compared to the O2 run in 2017, the main configuration change visible in Fig.~\ref{fig:Virgo_O3} (left) is the addition of a source of squeezed vacuum states of light in between the beam-splitter mirror (BS) and the interferometer output. During O3, the signal-recycling mirror location (SRM) hosted the first lens of the detection system. 
A specific working point is required to have AdV sensitive to GW. When one such signal passes through the detector, it has a differential effect on the arm optical paths, changing the interference condition at the output. The variation of the detected power is the main input (along various control and monitoring channels) used to reconstruct the GW strain channel $h(t)$ from the raw data. Many active feedback control systems bring the detector to its working point and maintain it. The sensitivity of AdV is limited by many noises, which can be classified into three main categories: fundamental noises (seismic, thermal or quantum), technical noises (from the detector control system or some parts of the apparatus), or environmental noises. Fighting against them is a continuous struggle which takes place at all steps of the operation cycle followed by a detector like AdV: design, construction, installation, commissioning and optimization, data taking, design again, etc. There is literally a phase of "noise hunting", with the noise sources identified during that stage being either fixed or mitigated.

\section{Performance of the Virgo detector during the LIGO-Virgo O3 run}

\begin{figure}[!htbp]
    \begin{center}
      \includegraphics[width=0.98\columnwidth]{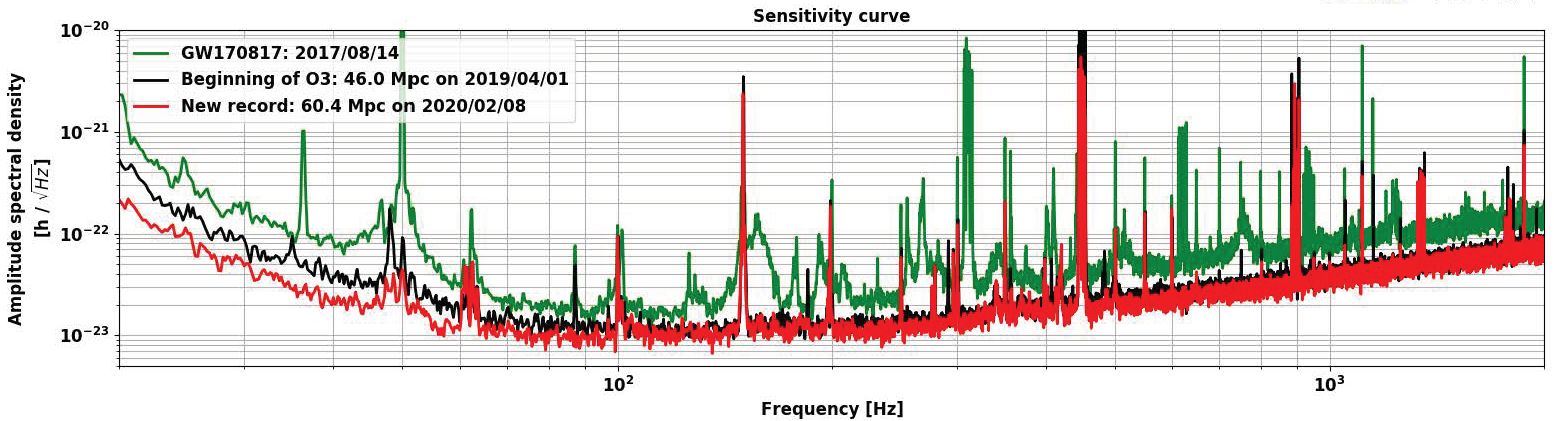}
    \end{center}
    \caption{AdV sensitivity, 2017-2020: O2 (green trace), beginning of O3a (black) and O3b record (red).}
    \label{fig:Virgo_sensitivity}
\end{figure}

\begin{table}[htbp!]
\centering
\begin{tabular}{ccccc}
\hline Category & Science & Activities & Control & Problems \\
\hline Percentage & 76.0 & 9.9 & 7.3 & 6.8 \\
\hline
\end{tabular}
\caption{\label{table:O3_Virgo_duty_cycle}Summary of the AdV O3 duty cycle.}
\end{table}
Table~\ref{table:O3_Virgo_duty_cycle}~\cite{2210.15633} shows the breakout of the O3 run duration into four different categories. The duty cycle ('Science' data-taking) is 76\%, with almost no impact of the season: ${\sim} 1\%$ difference between Spring-Summer and Fall-Winter. Then, about 10\% of the time is spent doing activities around the detector (commissioning, calibration and maintenance), 7\% working on bringing AdV to its working point and 7\% dealing with problems of various kinds. Projecting this duty cycle onto a baseline week (or a baseline day) by averaging all O3 data shows that its variations are mainly due to the activities of the detector crew described above. Removing them makes the duty cycle much more constant over the projected time range.
Compared to 2017, the AdV sensitivity has about doubled, as shown in Fig.~\ref{fig:Virgo_sensitivity}. Using the binary neutron star system (BNS) range\footnote{The sky location- and source orientation-averaged detection up to which the merger of a BNS can be detected with a signal-to-noise ratio of at least 8.} as a summary figure-of-merit, that number increases from 28~Mpc (2017) to 50+ and even 60~Mpc (O3b record).

\section{Environmental noises and their impact on the Virgo detector during the O3 run}

Environmental noises are monitored by hundreds of probes. Their location is shown in Fig.~\ref{fig:Virgo_O3} (right). The insert at the top shows a zoom around the central area (CEB: CEntral Building; MCB: Mode-Cleaner building) which contains most of the detector components, except for the two equivalent end buildings (NEB: North-End Building; WEB: West-End Building) located three kilometers apart. 
To fight environmental noises, all critical optical components (laser, mirrors, optical benches and sensors) are suspended to isolate them from seismic motion. These suspensions work well above a few Hz. Moreover, most of the AdV hardware is under high-vacuum: to avoid interaction between laser beams and air molecules, to keep optics clean, and to benefit from an acoustic shield. Additionally, all components designed, built or selected, must be low-noise. Tools like frequency band-limited RMS 
are used to extract some parts of interest from a given signal. For instance, these allows disentangling the different contributions to the seismic noise: microseism (up to 1~Hz), anthropogenic (from 1 to 5-10~Hz) and onsite (larger frequencies).

\begin{figure}[!htbp]
    \begin{center}
      \includegraphics[width=0.85\columnwidth]{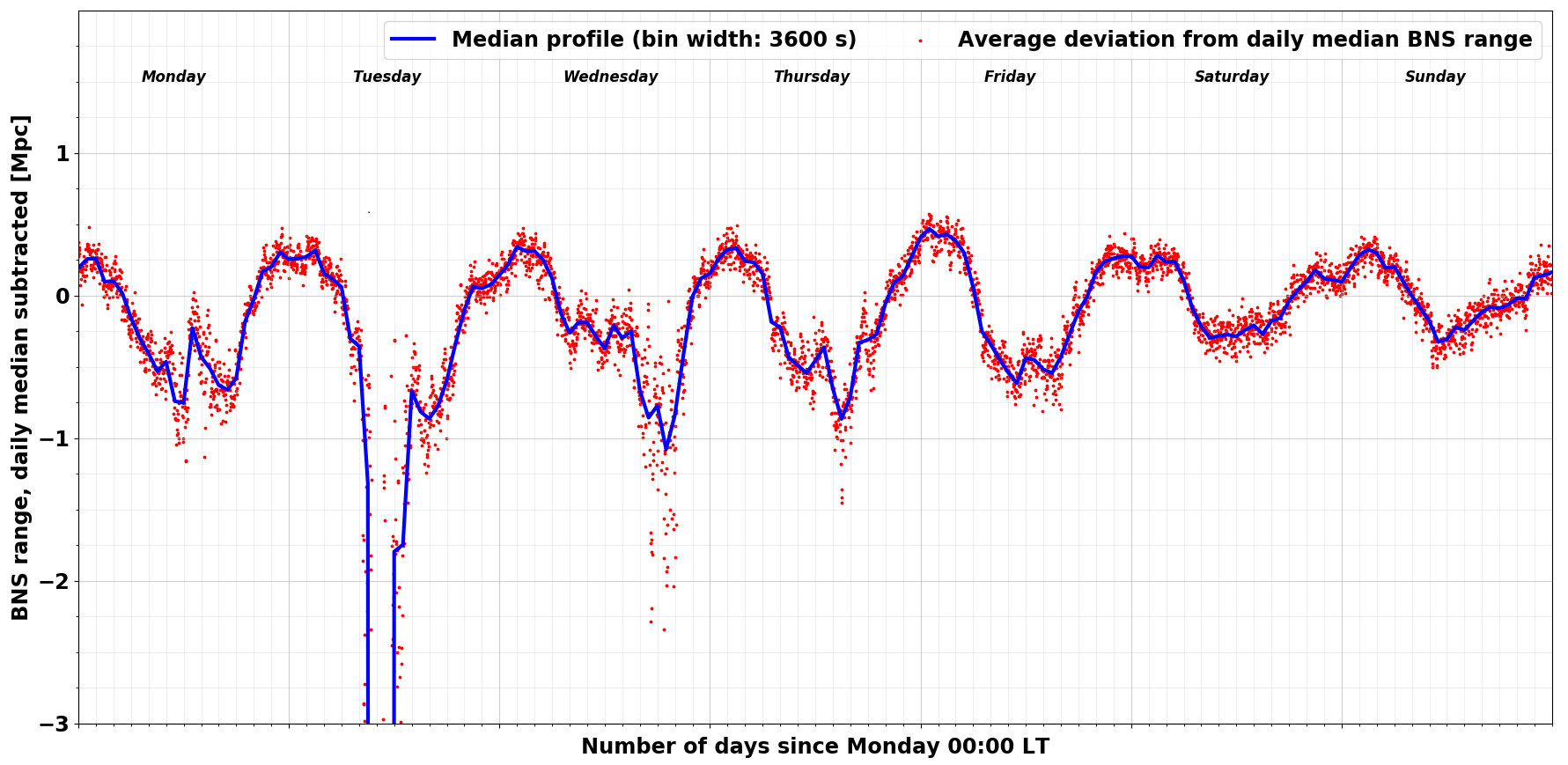}
    \end{center}
    \caption{O3-averaged fluctuations of the AdV BNS range as a function of the week day and time.}
    \label{fig:O3_virgo_range_robustness}
\end{figure}

\begin{figure}[!htbp]
    \begin{center}
      \includegraphics[width=0.8\columnwidth]{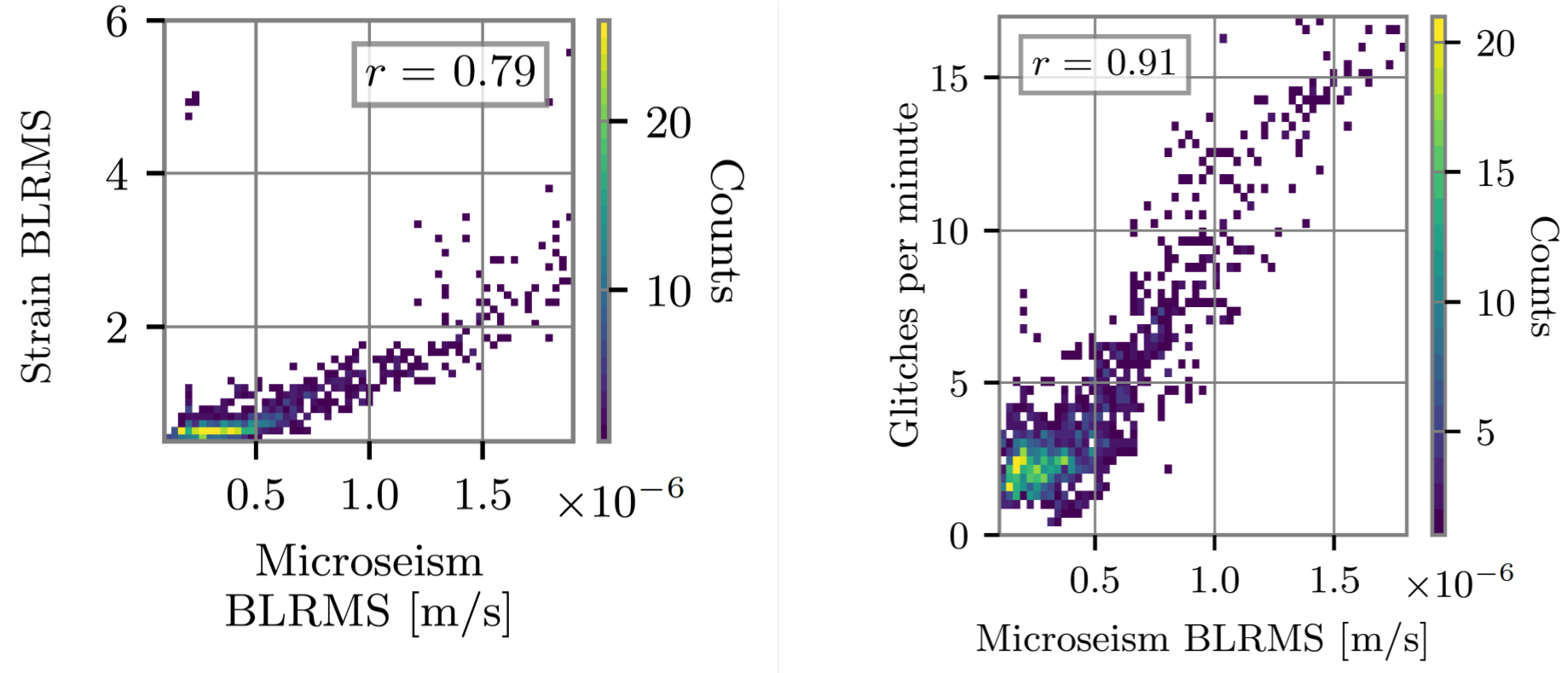}
    \end{center}
    \caption{Impact of microseism on the AdV noise.}
    \label{fig:microseism}
\end{figure}

Figure~\ref{fig:O3_virgo_range_robustness}~\cite{2203.04014} summarizes the modulation of the AdV sensitivity by projecting the BNS range variations around its daily median level over a week baseline. The observed changes follow those of the on-site anthropogenic noise: day-night variations and differences between week days and weekend. Their amplitude is small compared to the scale of the BNS range during the run, meaning that the detector was quite robust against this disturbance.
Microseism can impact the AdV noise in two ways, as shown on Fig.~\ref{fig:microseism}~\cite{2210.15633}. On the one hand, increasing the noise level at low frequency (10-20~Hz, left correlation plot, corresponding to O3b). A comparison between O3a and O3b shows that the noise has improved with time, with its residual level mainly due to microseism at the end of the run. On the other hand, by increasing the rate of transient noise bursts, called glitches (right plot, full O3 run).
Scattered light noise 
generate glitches or leads to control inaccuracies. Enhanced by high-microseismicity, they are one of the main technical noise sources for all GW detectors. Therefore, software tools aiming at locating scattering surfaces will run daily in the future.

\begin{figure}[!htbp]
    \begin{center}
      \includegraphics[width=0.49\columnwidth]{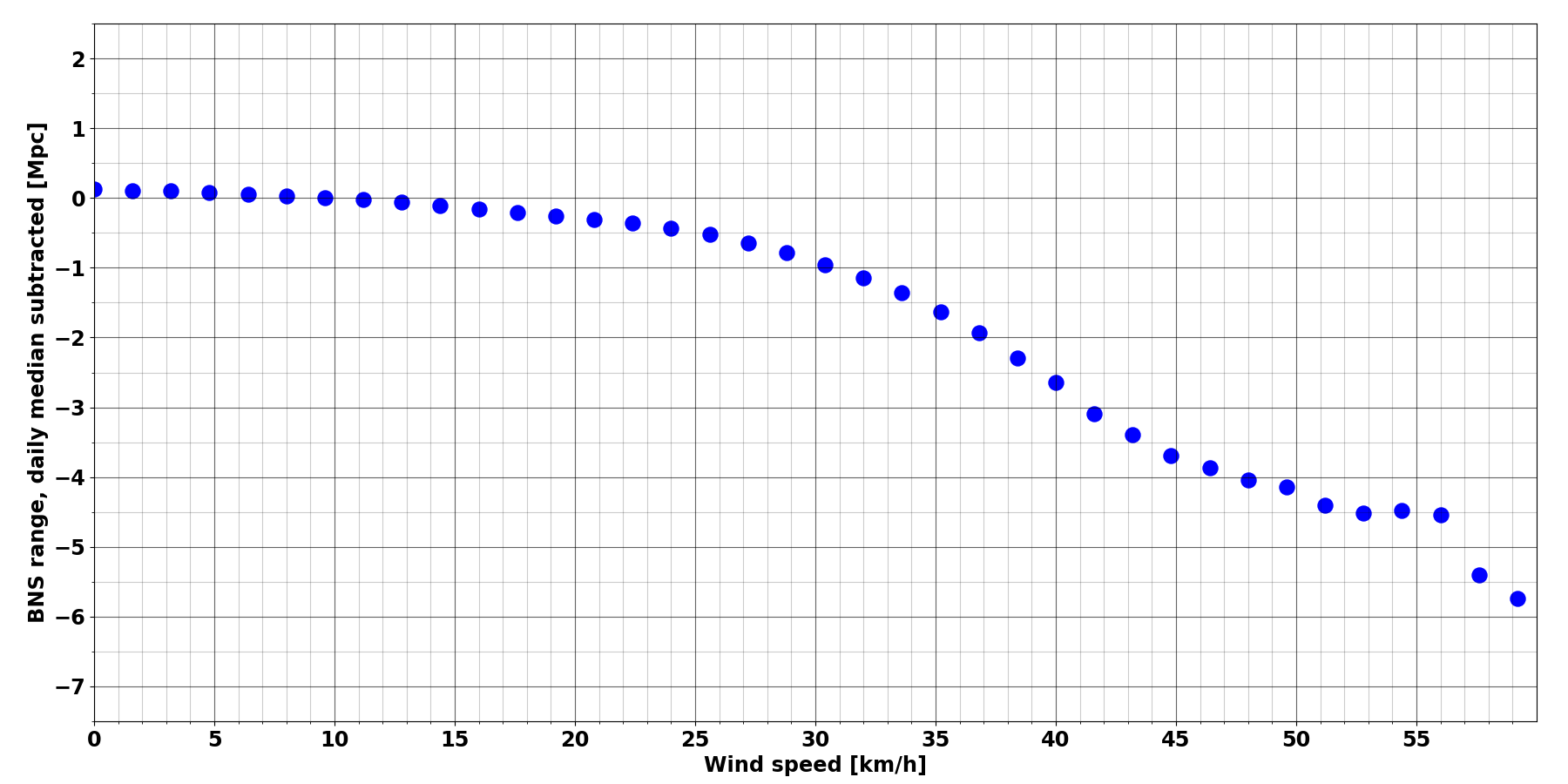}
      \includegraphics[width=0.49\columnwidth]{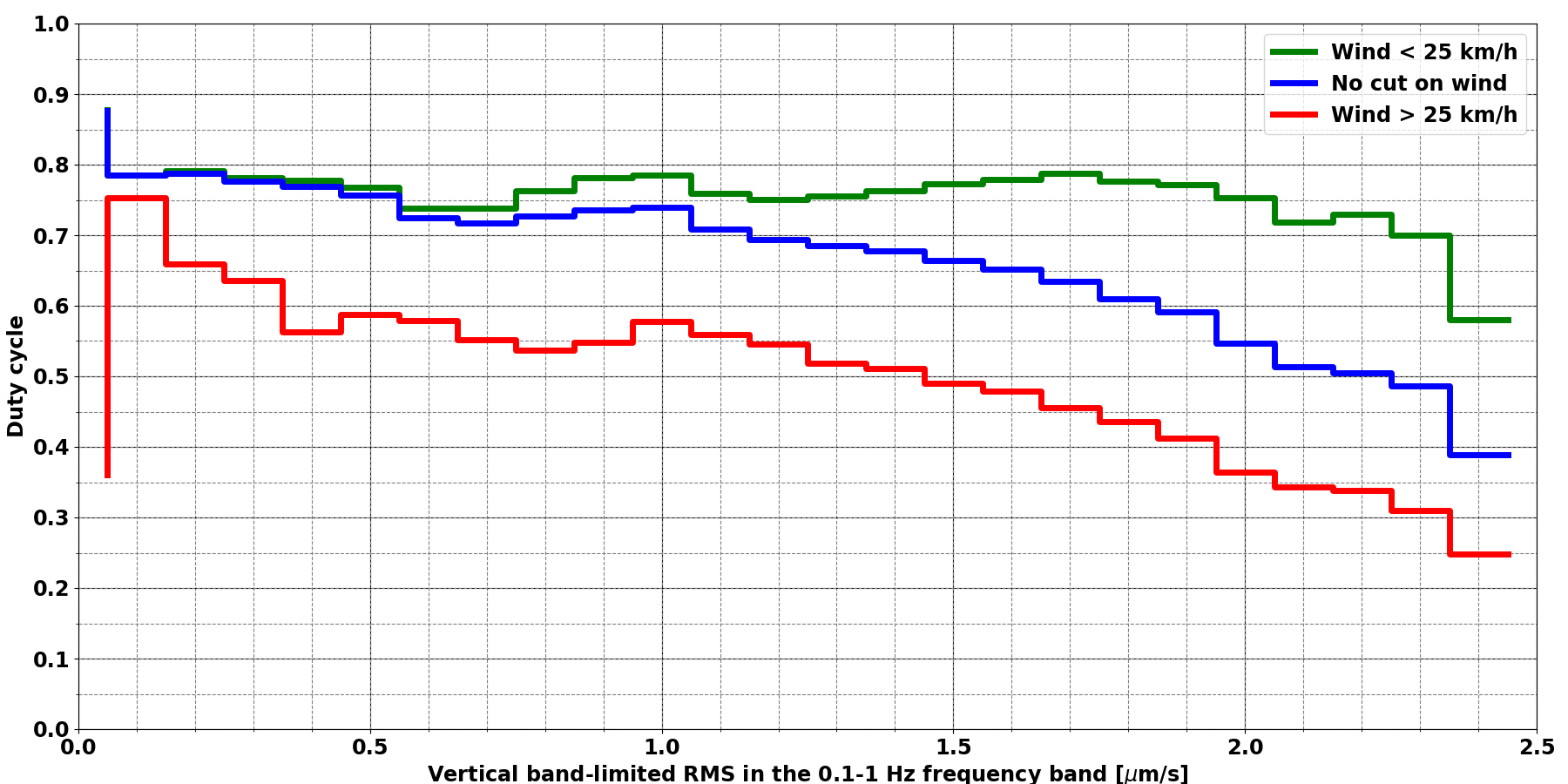}
    \end{center}
    \caption{Left: average variation of the AdV BNS range during O3, versus wind speed. Right: average AdV duty cycle versus microseism, for three different wind conditions: no cut (blue trace), low wind speeds (up to 25~km/h, green trace), high wind speeds (above 25~km/h, red trace). }
    \label{fig:wind_sea_activity}
\end{figure}

Bad weather is another source of elevated seismic, potentially impacting the detector in many ways: degraded sensitivity, reduced efficiency to detect transient GW signals, lower duty cycle, and more difficulties to optimize the instrument working point. Figure~\ref{fig:wind_sea_activity}~\cite{2203.04014} presents the result of a study aiming at disentangling the contributions from wind and rough sea. The left plot shows that the wind has little influence on the BNS range until a speed of ${\sim} 30$~km/h, while there is some larger degradation for higher wind speeds. The right plot shows that the detector is more resilient towards high microseism when the wind is low or moderate, meaning that AdV is more sensitive to the latter. More weather stations have been installed to get a better mapping of wind gusts in O4.

\begin{figure}[!htbp]
    \begin{center}
      \includegraphics[width=0.49\columnwidth]{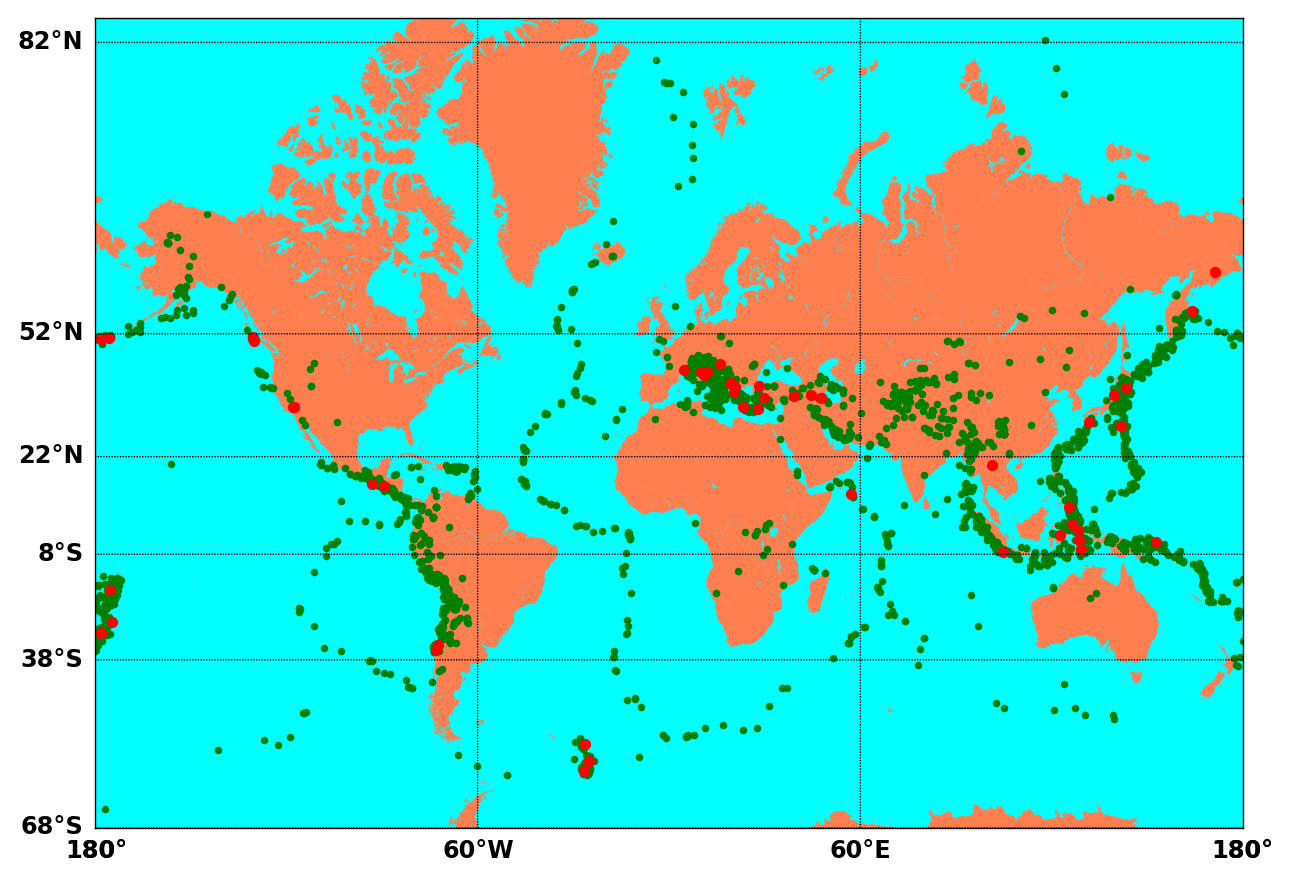}
      \includegraphics[width=0.49\columnwidth]{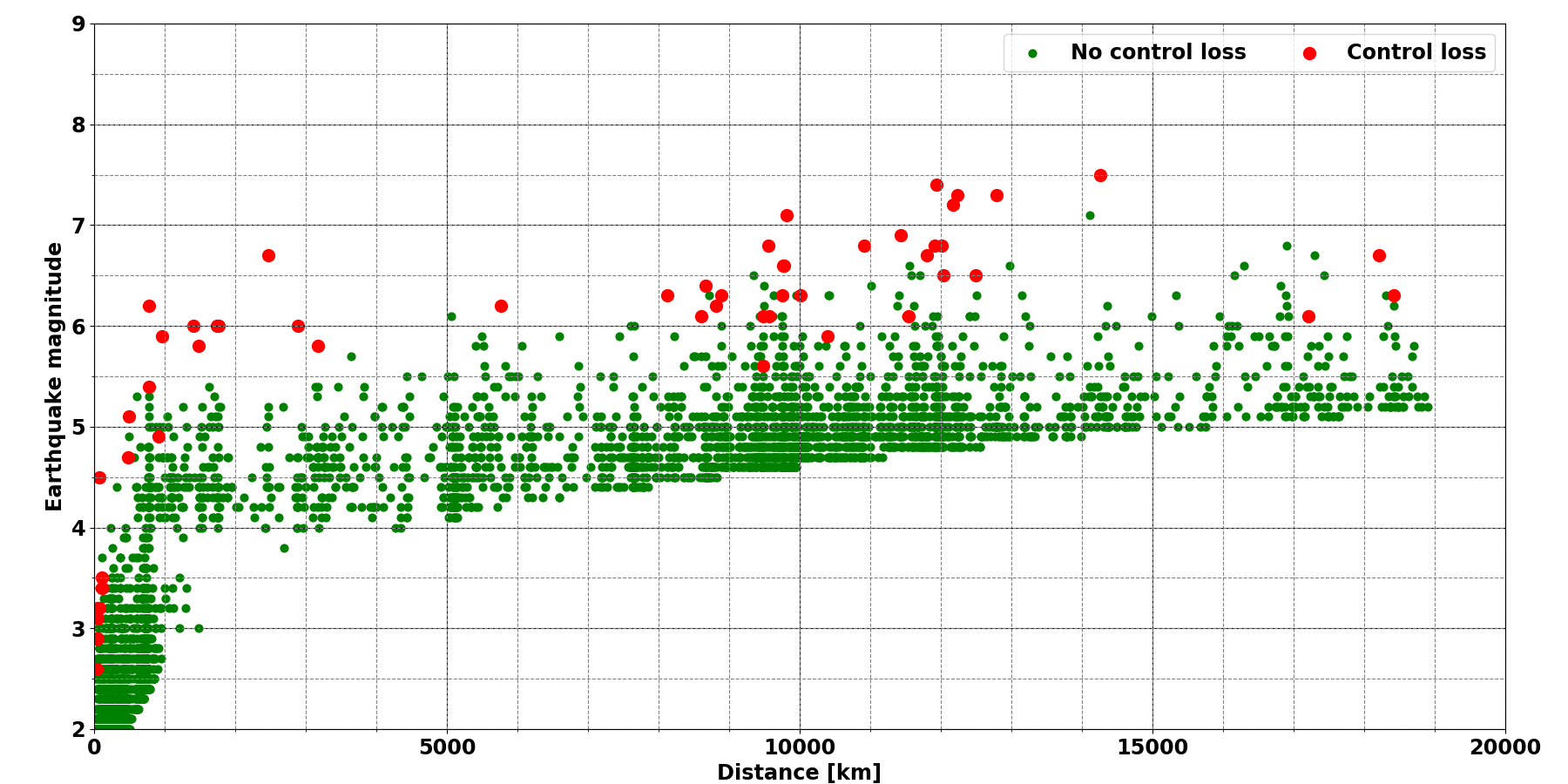}
    \end{center}
    \caption{Left: map of all 
O3 earthquakes
; the dot color says whether (green) or not (red) they triggered a detector control loss. Right: the same earthquakes, in the magnitude versus epicenter distance plane.}
    \label{fig:earthquakes}
\end{figure}

Earthquakes generate seismic waves which propagate from their epicenter at a few km/s. Although they get attenuated along the way, earthquakes which are large- or close-enough can make the AdV global feedback system saturate, leading to a loss of the working point control and thus causing a decrease of the duty cycle. Figure~\ref{fig:earthquakes}~\cite{2203.04014} shows the classification of the earthquakes recorded during O3 in two categories, depending on whether or not they triggered a loss of the control of AdV~-- harmless weak earthquakes have been removed. Early warning systems are being developed and used to get notified in advance of the arrival of seismic waves from distant earthquakes. The idea is to use the available time to assess the strength of these waves, and to decide whether some mitigation action should be attempted, with the goal of maintaining the detector control.

Last but not least, ambient magnetic fields can couple through the coil-magnet actuators. Moreover, given that electromagnetic waves propagate at the speed of light like the GWs, they could impact the network with time delays compatible with a real signal. Two examples of such phenomena are the Schuman resonances and large-current lightning strikes. Magnetic noise also includes an anthropogenic component. 
Electromagnetic devices outside the site and thus independent from AdV can also lead to huge magnetic noises, making compulsory the need of a constant monitoring, provided by a couple of external magnetometers.

\section{Conclusions}

O3 has been a great success for AdV: the detector has been online during the whole run and its duty cycle has been high and consistent, with a sensitivity significantly improved. Moreover, AdV appears to have been robust overall, against the external environment. The studies reported here will continue for O4 and beyond, as the instrument's global working point is complex and changes significantly after each hardware upgrade. The AdV monitoring will be improved by using tools which will be more automated, provide results in lower latency and have larger scopes.

\bibliographystyle{JHEP}
\bibliography{proceedings_ICHEP2022_NA.bib}

\end{document}